\documentclass{article}
\usepackage{amssymb}
\usepackage{amsmath}

\input{tcilatex}

\begin{document}

\title{The Orthopositronium Problem and $e$-$\mu $-$\tau $-Universality}
\author{B.M.Levin$^{\ast }$ \\
\textit{A.F.Ioffe Physical Technical Institute, }\\
\textit{\ 194021 St. Petersburg, Russia}}
\maketitle

\begin{abstract}
The quantitative description of the orthopositronium anomalies
(\textquotedblleft isotope anomaly\textquotedblright\ in a gaseous neon for
the \textquotedblleft resonance conditions\textquotedblright\ and
\textquotedblleft $\lambda _{T}$-anomaly\textquotedblright\ in non-resonance
conditions) is possible on the basis of a hypothesis about restoration of
spontaneously broken \textit{complete relativity} (A.F.Andreev, 1982) of the
limited space-time \textquotedblleft volume\textquotedblright\
(\textquotedblleft defect\textquotedblright\ of the space-time) in a final
state of $\beta ^{+}$-decay ($\Delta J^{\pi }=1^{+}$) of nucleus such as $%
^{22}Na$, $^{68}Ga$, etc. This addition of the Standard Model in a final
state of the topological quantum transition (non-steady-state, a generalized
\textquotedblleft displacement current\textquotedblright ) supposes the
description on the common basis timelike fundamental particles (locality)
and spacelike fundamental structure (\textit{non-locality without causal
paradoxes}). One of achievements of expansion of Standard Model can be a
substantiation of topological connection \textit{e}-$\mu $-$\tau $%
-Universality with discrete structure (quantization) \textquotedblleft
defect\textquotedblright\ of the space-time.
\end{abstract}

Consideration by analogy with orthopositronium [$o-Ps\ \equiv \
^{3}(e_{\beta }^{+}\ e^{-})_{1}$] of the atoms $^{\ 3}(\mu ^{+}\mu
^{-})_{1}\ $and$\ ^{3}(\tau ^{+}\tau ^{-})_{1\ }$ finds out basic difference
of the top generations of leptons: impossibility of realization for $^{\
3}(\mu ^{+}\mu ^{-})_{1}^{\ }$-$\ $and\ $^{3}(\tau ^{+}\tau ^{-})_{1}$%
-atoms, even in some Gedanken experiment, a conditions similar to the
\textquotedblleft resonance conditions\textquotedblright\ at supervision of
the \textquotedblleft isotopic anomaly\textquotedblright\ in samples of a
gaseous neon of various isotope abundance [1,2]. If to treat it as
spontaneous breaking of the horizontal symmetry of the leptons generations,
then their participation in formation of the fundamental space-like
structure within the framework of \textquotedblleft \textit{additional} $%
G\hbar /c$-\textit{physicis}\textquotedblright\ would be natural consequence
of restoration of the mentioned symmetry in a final state of $\beta ^{+}$%
-decay for a time $\tau _{\mu }\sim R_{\mu }/c\sim 10^{-6}$ $s$\ [3-5].

Let's present discrete, space-like and two-sign ($\pm $, on mass and all
charges, including baryon charge) structure as a tachyonic medium [6],
formed in a final state of the $\beta ^{+}$-decay of nucleus $^{22}Na$, etc $%
(\Delta J^{\pi }=1^{+})$. The linear sizes of the fundamental space-like
structure presented in [4] -- \textquotedblleft virtual fundamental
length\textquotedblright\ $\Delta \simeq 5.5\cdot 10^{-2}$ $cm$ and the size
of the \textquotedblleft long-range atom\textquotedblright\ $2R_{\mu }\simeq
1.1\cdot 10^{5}$\ $cm$ -- we shall compare with a the characteristic length
of the tachyonic instability [6]%
\begin{equation}
L>C\ /\ \Gamma ,  \tag{1}
\end{equation}%
where $C\ $-- the characteristic rate agreed in this specific case with
velocity of light$\ c$, and $\Gamma $ is an inverse time of development of
instability ($s^{-1}$).

For revealing the physical content of the assumed comparison is important
presented in [6] isomorphism of the equations of the tachyonic field 
\begin{equation*}
\left( \omega ^{2}+C^{2}\cdot \nabla ^{2}+\Gamma ^{2}\right) \cdot \Psi =0\
\ \ \ \ \ (\hbar =1)
\end{equation*}%
and of the Schr\"{o}dinger equation for the stationary state of the
\textquotedblleft particle\textquotedblright\ with mass $m$ $(m=1/2)$%
\begin{equation*}
\left[ \nabla ^{2}+E-V(x)\right] \cdot \Psi =0,
\end{equation*}%
i.e. a conformity takes place%
\begin{equation*}
\omega ^{2}/C^{2}\iff E,\ \ \ \ \Gamma ^{2}/C^{2}\iff -V(x),
\end{equation*}%
if to consider value $\Gamma $ depending from the spatial coordinate $x$.

It follows that tachyonic instability $(\omega ^{2}<0,\ \omega =i\Gamma ,\
\Psi \sim (exp\left\vert \Omega \right\vert t)$ is appropriate to the bound
state $-\Gamma ^{2}/C^{2}\ $in an attraction field , and a negative energy
of a compensating field (the \textquotedblleft \textit{mirror Universe}%
\textquotedblright\ [3-5]) can be presented as the binding energy $%
E_{b}=\left\vert M_{Pl}\right\vert \cdot c^{2}\ $of the vacuumlike substance
component with mass $M_{Pl}>0\ $of the whole fundamental space-like
structure. We receive the increment increase of the tachionic field [6]%
\begin{equation*}
\Psi \sim exp(\Omega t),\ \ \ \Omega =C\cdot (-E_{b})^{1/2},
\end{equation*}%
and $\Omega $ in a considered context accepts value%
\begin{equation}
\Omega =\frac{c\cdot \lbrack 2M_{Pl}(-M_{Pl}\cdot c^{2})]^{1/2}}{\hbar }\sim
it_{Pl}^{-1}\sim i\cdot 10^{43}\ s^{-1}.  \tag{2}
\end{equation}

All \textquotedblleft sites\textquotedblright\ of the crossing of
\textquotedblleft spaces\textquotedblright\ $X\ $(in the observable world)
and $X^{^{\prime }}$ (an inverse \textquotedblleft space\textquotedblright\
-- the \textquotedblleft mirror Universe\textquotedblright )
\textquotedblleft stick together\textquotedblright\ in \textquotedblleft
point\textquotedblright\ in the space $3+N^{^{^{\prime }}}$-dimensions: $%
\{X\}=\{X^{^{\prime }}\}=0$. In the space of the observer it allocates the
"central" site of "lattice" of the fundamental macroscopical space-like
structure with which in a final state of the $\beta ^{+}$-decay $^{22}Na\
(e^{+},$ $\nu )^{22\ast }Ne$ the daughter nucleus $^{22\ast }Ne\ $has been
linked. The four-dimensional fundamental space-like structure of the size $%
2R_{\mu }$ in the $3+N^{^{\prime }}$-dim space is captured by Planckian
length $l_{Pl}$. Process of the tachyonic instability by the conserve of the
angular momentum is limited. Specifically, a \textquotedblleft
self-unwinding\textquotedblright\ of the fundamental space-like structure in
the $\beta ^{+}$-decay final state by a difference of the angular moments of
the final and initial states ($^{22}Na,\ ^{68}Ga,\ $etc) is limited%
\begin{equation}
\Delta J=\hbar >I_{\mu }\cdot c/R_{\mu }\ ,  \tag{3}
\end{equation}%
as velocity of the fundamental space-like structure on its \textquotedblleft
surface\textquotedblright\ cannot exceed the velocity of light. In (3) $%
I_{\mu }\sim m_{\mu }\cdot R_{\mu }^{2}\ $is the moment inertia of a
spherical huming-top, $m_{\mu }$ $\sim \hbar /R_{\mu }c\ $is its mass.
Hence, practically instant "self-unwinding" of the fundamental space-like
structure ($\left\vert \Omega \right\vert \sim 10^{43}\ s^{-1}$)
periodically undergoes a failure and again renews with a casual directions
of the angular moment. This result give proof to the postulate about the
casual rotation of the \textquotedblleft mirror Universe\textquotedblright\
with a velocity $V\ \sim -\ c$ in respect to the ground-based laboratory (to
an \textit{observer}) [7].

The observer perceives additional discrete structure of the
\textquotedblleft mirror Universe\textquotedblright\ as discrete scalar 
\textit{C-field} [8] with a negative sign on all physical \textquotedblleft
charges\textquotedblright\ (including mass of "holes" -- "proton" and
"electron" in the structure of "lattices") [4]. \textit{C}-field compensates
of the limited \textquotedblleft volume\textquotedblright\ of the space-time
($\sim 1$ $km^{3}$, during $\sim 2\cdot 10^{-6}$ $s$) -- the \textit{%
vacuumlike state of matter} (VSM) [9] ( the \textquotedblleft \textit{%
long-range atom}\textquotedblright\ with \textit{Planckian mass} also
perceived by the observer as microstructure of the VSM and $1$-dim discrete
space ($N^{^{\prime }}=1$).

As a whole the final state of the $\beta ^{+}$-decay ($\Delta J^{\pi }=1^{+}$%
) represents 5-dim ($3+1^{^{\prime }}$, $t$) with discrete structure of the
spatial dimensions in the limited \textquotedblleft
volume\textquotedblright\ of the space-time.

Inequalities (comp. with (1) )%
\begin{equation}
\Delta >\frac{c}{\Gamma _{\tau }}\ cm\text{ \ \ and \ \ }2R_{\mu }>\frac{c}{%
\Gamma _{\mu }}\ cm  \tag{4}
\end{equation}%
where $\Gamma _{\tau }\ $($\sim 3.3\cdot 10^{12}\ s^{-1}$)\ and $\Gamma
_{\mu }\ $($\sim 4.5\cdot 10^{5}\ s^{-1}$) , accordingly, are the decay
width of the $\tau $-$\ $and $\mu $-leptons, show that in structure of the
value $\Gamma (x)\ $is available the contribution of a topological component
of the increase of the tachyonic field -- width of decay of leptons $\Gamma
_{\tau },\ \Gamma _{\mu }$ ($\Gamma _{\tau \ /\ \mu }$), defining the rate
development of an instability.

The topological component limits the rate of increase of the instability as
a time of the consecutive transitions is defined by sum of characteristic
times%
\begin{equation*}
\frac{1}{\Gamma \left( x\sim \Delta \ /\ x\sim R_{\mu }\right) }=\frac{1}{%
\Gamma _{\tau \ /\ \mu }}+\frac{1}{\mid \Omega \mid }
\end{equation*}%
and%
\begin{equation}
\Gamma \left( x\sim \Delta \ /\ x\sim R_{\mu }\right) =\frac{\Gamma _{\tau \
/\ \mu }\cdot \mid \Omega \mid }{\Gamma _{\tau \ /\ \mu }+\mid \Omega \mid }%
\cong \Gamma _{\tau \ /\ \mu }\ \ \ \ (\mid \Omega \mid >>\Gamma _{\tau /\mu
}).  \tag{5}
\end{equation}

\bigskip

The width of the leptons decay is defined by the the weak interaction
constant. Thus, unification of the all interactions on the basis of
antipodic symmetry [10] in the \textquotedblleft additional $G\hbar /c$%
-physics\textquotedblright\ (\textit{superantipodic symmetry }[3]) on the
basis of their own coupling constants take place, instead of a uniform
constant as it is supposed in the standard \textquotedblleft Theory of
Everything\textquotedblright\ at ultrahigh energy. It means, that the
impossibility to fix a \textit{coupling constant} for the \textit{Goldstone
field} in the theory with spontaneously broken complete relativity [11] is
not lack of the theory.

In the offered concept \textquotedblleft additional $G\hbar /c$%
-physics\textquotedblright\ the top generations of the leptons adjust the
bifurcation attitude mutually supplementing aspects of strong interaction:
one -- with participation of color (quantum cromodynamics/aromadynamics)
[12], and another -- a non-stationary long-range for a baryon (and lepton)
charge [3-5].

Let's note, that the estimation of the mass presented here fundamental
spacelike structures $m_{\mu }$ $\sim \hbar /R_{\mu }c~\sim 3\cdot 10^{-10}eV
$ is kept at extremely wide limits admissible by Standard Model for mass
invisible acsion, is closer to its lower limit $\sim 10^{-12}~eV$. If the
\textquotedblleft long-range atom\textquotedblright\ at the moment of a
birth in final state of $\beta ^{+}$-decay or as result of the its
subsequent diffusion is brought into gravitational field with the critical
value of acceleration of free falling%
\begin{equation*}
\gamma _{cr}=\frac{\hbar \cdot c}{m_{p}\cdot R_{\mu }^{2}}\geqslant 0.01\
cm/s^{2},
\end{equation*}%
there is its \textquotedblleft splitting\textquotedblright\ on plus%
\TEXTsymbol{\backslash}minus Planck masses (non-steady-state, the \textit{%
generalized displacement current}), i.e. in one-stage (from
\textquotedblleft \textit{nothing}\textquotedblright !) \textquotedblleft
elementary\textquotedblright\ macroscopical \textquotedblleft
domain\textquotedblright\ of a dark matter with mass 2$\left\vert
M_{Pl}\right\vert $\ is born [13].

$^{\ast }$E-mail: bormikhlev@mail.ioffe.ru, bormikhlev@mail.ru

\end{document}